\documentclass[
aps,pra,
 ,twocolumn
]{revtex4}
\usepackage{graphicx}
\usepackage{amsmath}
\usepackage{amssymb}
\usepackage{amsthm}
\usepackage{color}
\usepackage{hyperref}
\hypersetup{colorlinks,linkcolor={blue},citecolor={red},urlcolor={blue}}
\usepackage{tikz}

\newcommand{\ket}[1]{| #1 \rangle}
\newcommand{\bra}[1]{\langle #1 |}

\newcommand{\vt}[1]{{\boldsymbol{#1}}}
\theoremstyle{plain}
\newtheorem{thm}{Theorem}
\theoremstyle{plain}

\newtheorem{conjecture}[thm]{Conjecture}

\begin{document}

\title{Generation of photonic non-Gaussian states by measuring multimode Gaussian states}

\author{Daiqin Su}
\email{daiqin@xanadu.ai}
\author{Casey R.~Myers}
\author{Krishna Kumar Sabapathy}
\affiliation{Xanadu, 777 Bay Street, Toronto, Ontario, M5G 2C8, Canada}


\date{\today}

%
\begin{abstract}
We present a detailed analytic framework for studying multimode non-Gaussian states that are conditionally generated when few modes of a multimode Gaussian state are subject to photon-number-resolving detectors. 
From the output state Wigner function, we deduce that the state factorizes into a  Gaussian gate applied to a finite Fock-superposition non-Gaussian state. 
The framework provides an approach to find the optimal strategy to generate a given target non-Gaussian state. We explore examples, such as the generation of cat states, weak cubic phase  states, and bosonic code states, and achieve improvements of success probability over other schemes. 
Our framework also applies to the case in which the measured Gaussian state is mixed which is very important for the analysis of experimental imperfections such as photon loss. The framework has potential far-reaching implications to the generation of bosonic error-correcting codes and for the implementation of non-Gaussian gates using  resource states, among other applications requiring non-Gaussianity.  
\end{abstract}

\maketitle

{\it {Introduction}.}  --  Non-Gaussian states and non-Gaussian gates are crucial and essential ingredients in quantum information processing and universal quantum computation using continuous-variable systems \cite{RevModPhys.84.621,RevModPhys.77.513}. 
However, generating non-Gaussian states in a deterministic manner remains a challenge in  quantum optics due to  weak interaction Hamiltonians that are polynomials in the quadrature operators with order $>2$, such as the Kerr interaction. An alternative is to herald non-Gaussian states through photon-number measurements,  such as photon subtraction \cite{PhysRevA.55.3184}.  Photon 
subtraction has been used to generate non-Gaussian states like Schr\"odinger's cat states \cite{PhysRevA.55.3184, ourjoumtsev2006generating, PhysRevLett.97.083604, PhysRevLett.101.233605, PhysRevA.82.031802}, NOON states \cite{PhysRevA.40.2417, PhysRevLett.85.2733}, superpositions of Fock states \cite{yukawa2013,fiuravsek2005conditional}, photonic tensor network states \cite{PhysRevLett.120.130501}, error-correcting bosonic code states \cite{PhysRevA.56.1114, PhysRevA.94.012311, PhysRevA.97.032346}, and to tailor more complicated Gaussian states like continuous-variable cluster states \cite{PhysRevLett.121.220501}.

An important challenge using photon subtraction is that the success probability is low for engineering complicated target states.  
So we set out the task to use minimal resources of squeezed displaced vacuum states, interferometers, and photon-number-resolving detectors to find optimal circuits for given target states. 
Recently, a machine learning method was used to search for such circuits 
that resulted in an improved success probability of four orders of magnitude for the generation of weak cubic phase states with near-perfect fidelity \cite{sabapathy2018near}. Furthermore, from an experimental point of view  photon-number-resolving detectors (PNRDs)  are now readily available to use for the generation of
multiphoton states \cite{magana2019multiphoton, tiedau2019scalability}.

\begin{figure}
\includegraphics[width=\columnwidth]{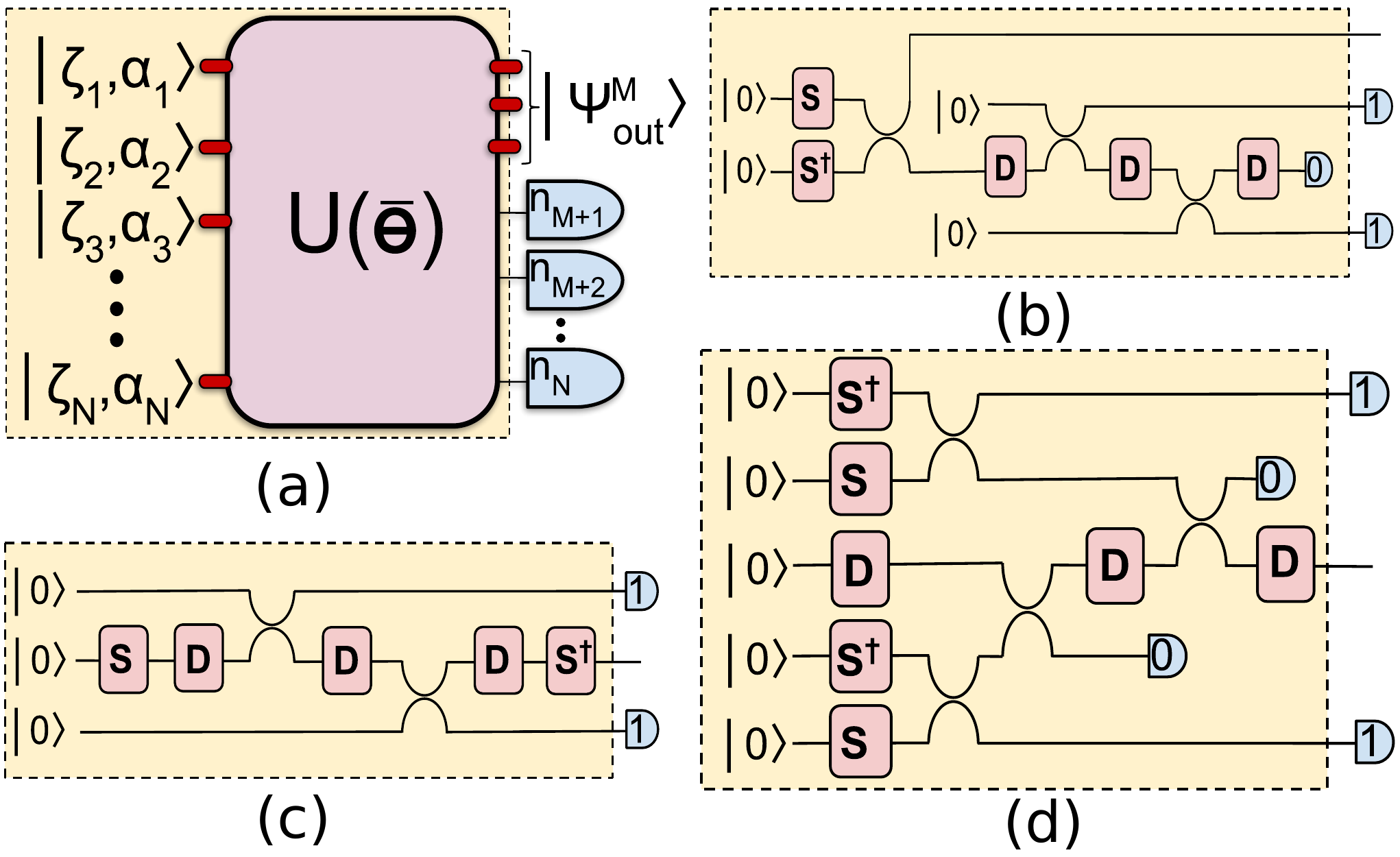}
\caption{Optical schemes for the generation of non-Gaussian states. (a) Our method to measure few modes of a multimode pure Gaussian state ($\ket{\zeta_i,\alpha_i}$ is a squeezed displaced vacuum state in the $i^{th}$  mode with squeezing $\zeta_i$ and displacement $\alpha_i$, $U(\bar{\theta})$ is an interferometer, $n_j$ are photon-number-resolving-detector (PNRD) outcomes). (b) Application of repeated displacements and photon subtractions to one arm of a two-mode squeezed vacuum state \cite{yukawa2013}. (c) Utilization of repeated photon subtractions and displacements on a squeezed vacuum state \cite{fiuravsek2005conditional}. (d) Application of repeated displacements and photon additions \cite{dakna1999generation}. The dashed regions in (b)-(d) can be mapped to a particular instance of the dashed region in (a). Thus our scheme is the most general heralding scheme using input pure Gaussian states and photon-number-resolving (PNR) measurements.  }
\label{fig:unifiedschemes}
\end{figure}

In this paper, we develop a general framework to study the generation of non-Gaussian states by measuring an arbitrary multimode Gaussian state using 
PNRDs. Consider an arbitrary $N$-mode Gaussian state and measure $(N-M)$ modes using 
PNRDs, and postselect a certain measurement pattern, resulting in an $M$-mode output non-Gaussian state. 
This framework subsumes many previous state preparation schemes, as shown in Fig.~\ref{fig:unifiedschemes}. 
We derive an analytic formula for the conditional generation of general non-Gaussian state along with its success probability. 
The task of finding the optical circuit that gives the highest success probability for a given target state can also be obtained using our framework, although it is a more intricate procedure. 

{\it {Single-mode output states}.} -- We start from the simplest case where ($N-1$) modes of an $N$-mode pure Gaussian state are measured, resulting
in a single-mode non-Gaussian state. Generalizations to multimode outputs or measuring mixed Gaussian states is straightforward. 
Let us define an operator vector 
$\hat{\vt{\xi}}^{(c)} = (\hat a_1^{\dag}, \cdots, \hat a_N^{\dag}, \hat a_1, \cdots, \hat a_N)^{\top}$, where  $\hat a_k^{\dag} (\hat a_k)$ are the creation (annihilation) operators of the 
$k$-th optical mode that satisfy the boson commutation relation $[\hat a_j, \hat a_k^{\dag}] = \delta_{j k}$, the superscript ``(c)" represents the coherent state basis and we use bold symbols to signify vectors or matrices. Gaussian states are fully characterized by the mode operator's first and second 
moments, given explicitly as the displacement vector $\vt{Q}^{(c)} = \big\langle \hat{\vt{\xi}}^{(c)} \big\rangle$ and covariance matrix ${\bf V}^{(c)}$
\begin{eqnarray}
V^{(c)}_{jk} = \frac{1}{2} \big\langle \big\{ \hat{\xi}^{(c)}_j, \, \hat{\xi}^{(c) \dag}_k \big\} \big\rangle - \big \langle \hat{\xi}^{(c)}_j \big\rangle \big \langle \hat{\xi}^{(c) \dag}_k \big\rangle, 
\end{eqnarray}
respectively. Without loss of generality, we assume that the last $(N-1)$ modes are measured  onto the Fock state $\ket{\bar{\vt n}} = \ket{n_2, n_3, \cdots, n_N}$, where
$n_k$ is the photon number registered at the $k$-th PNRD. The output density matrix (unnormalized) of the first mode is $\tilde \rho_1 = \bra{\bar{\vt n}} \rho \ket{\bar{\vt n}}$ with a success probability $P(\bar{\vt n}) = \text{Tr}(\tilde \rho_1)$, where $\rho$ is the density matrix of the $N$-mode Gaussian state. 
The Wigner function of $\tilde \rho_1$ can be derived as \cite{su2019conversion}
\begin{align}\label{eq:WignerPure}
&W(\vt{\alpha};\tilde{\rho}_1) 
\propto \exp\bigg\{- (\vt{\alpha}^{\dag} - \vt{d}^{\dag} ) \big({\bf S} {\bf S}^{\dag} \big)^{-1} (\vt{\alpha} - \vt{d}) \bigg\} \nonumber\\
&\times \prod_{k=2}^{N} \bigg(\frac{\partial^2}{ \partial \alpha_k \partial \beta_k^*} \bigg)^{n_k}
\exp\bigg( \frac{1}{2} \vt{\gamma}_d^{\top} {\bf A} \vt{\gamma}_d + \vt{z}^{\top} \vt{\gamma}_d \bigg) \bigg|_{\vt{\gamma}_d = 0},
\end{align}
where $\vt{\gamma}_d = (\beta_2^*, \beta_3^*, \cdots, \beta_N^*, \alpha_2, \alpha_3, \cdots, \alpha_N)^{\top}$ and
$\vt{\alpha} = (\alpha^*, \alpha)^{\top}$. 
The output state depends on ${\bf V}^{(c)}$ and
$\vt{Q}^{(c)}$ of the initial measured Gaussian state, along with the PNRD pattern. The relation between ${\bf S}$, $\vt{d}$, ${\bf A}$, $\vt{z}$ and 
${\bf V}^{(c)}$, $\vt{Q}^{(c)}$ can be developed as follows. From the covariance matrix ${\bf V}^{(c)}$ and displacement $\vt{Q}^{(c)}$ 
we define a matrix $\widetilde{\bf R}$ and a vector $\tilde{\vt y}$ as
\begin{eqnarray}
\widetilde{\bf R} &=& {\bf X}_{2N} \big( 2 {\bf V}^{(c)} -  {\bf I}_{2N} \big) \big( 2{\bf V}^{(c)} + {\bf I}_{2N} \big)^{-1}, \nonumber\\
\tilde{\vt{y}} &=& 2 \, {\bf X}_{2N} \big(  2{\bf V}^{(c)} + {\bf I}_{2N} \big)^{-1} \, \vt{Q}^{(c)}, 
\end{eqnarray}
where ${\bf I}_{2N}$ is a $2N \times 2N$ identity matrix and $ {\bf X}_{2N} = {\bf X}_2 \otimes {\bf I}_{N}$ with ${\bf X}_2 = \begin{pmatrix} 0 & 1 \\ 1 & 0 \end{pmatrix}$. 
When the input Gaussian state is pure, it can be shown~\cite{PhysRevLett.119.170501} that
$\widetilde{\bf R} = {\bf B} \oplus {\bf B}^*$, where $ {\bf B}$ is an $N \times N$ symmetric matrix (with entries $b_{ij}$) given by 
$ {\bf B} = {\bf U} \bigoplus_{j=1}^{N} \tanh(r_j) \, {\bf U}^{\top}$
with $r_j$ the squeezing parameters of the input squeezed states and ${\bf U}$ the unitary matrix representing the linear interferometer. Note that the phases of the initial squeezed states can be absorbed into the interferometer (Fig.~\ref{fig:unifiedschemes} a). 
A permutation matrix ${\bf P}$ which moves the $(N+1)$-th component
of $\tilde{\vt{y}}$ to the second component can be used to define a new vector $\vt{y} = {\bf P} \tilde{\vt{y}}$ and a new matrix ${\bf R} = {\bf P} \widetilde{\bf R} {\bf P}^{\top}$. 
It is then easy to divide the heralded part (denoted $h$) and detected part (denoted $d$) in ${\bf R}$ and ${\vt{y}}$ as
${\bf R} = \begin{pmatrix} {\bf R}_{hh} & {\bf R}_{hd} \\ {\bf R}_{dh} & {\bf R}_{dd} \end{pmatrix}$ and ${\vt{y}} = ({\bf y}_{h},  {\bf y}_{d})^{\top}$, respectively. 
Now  ${\bf S}$, $\vt{d}$, ${\bf A}$, $\vt{z}$ can be written as
\begin{eqnarray}
{\bf S} &=& \frac{{\bf I}_2 + {\bf X}_2 {\bf R}_{hh}}{\sqrt{1 - |b_{11}|^2}}, \nonumber\\
\vt{d} &=& ({\bf I}_2 - {\bf X}_2 {\bf R}_{hh})^{-1} {\bf X}_2 \vt{y}_h, \nonumber\\
{\bf A} &=& {\bf R}_{dd} - {\bf R}_{dh} ({\bf I}_2 + {\bf X}_2 {\bf R}_{hh} )^{-1} {\bf X}_2 {\bf R}_{hd}, \nonumber\\
\vt{z} &=& {\vt Y} + \frac{2}{\sqrt{1-|b_{11}|^2}} \, {\bf R}_{dh} \, {\bf S}^{-1} (\vt{\alpha} - \vt{d}), 
\end{eqnarray}
where ${\vt Y} = \vt{y}_d + {\bf R}_{dh} ({\bf I}_2 - {\bf X}_2 {\bf R}_{hh})^{-1} {\bf X}_2 \vt{y}_h$. Note that ${\bf S} \in Sp(2,\mathbb{C})$, the group of  complex $2 \times 2$ symplectic matrices.

The Wigner function in Eq.~\eqref{eq:WignerPure} factorizes into two parts, a Gaussian function followed by a  
polynomial in $\alpha$. This implies that the output state can be written as 
\begin{eqnarray}\label{eq:StateStructureSingle}
\ket{\psi_1} = \hat D(\beta) \hat S(\zeta) \sum_{n=0}^{n_{\text{max}}} c_n \ket{n},
\end{eqnarray}
which is a displaced and squeezed superposition of Fock states, also noticed in the special case considered in~\cite{PhysRevA.72.033822}. 
The squeezing amplitude $\zeta$ is determined by $b_{11}$:
$|\zeta| = \frac{1}{2} \ln \big( \frac{1 + |b_{11}|}{1 - |b_{11}|}\big)$ and $\text{arg} (\zeta) = -\text{arg} (b_{11})/2$. The  displacement $\beta$ is determined by 
$\vt{d} = (\beta^*, \beta)^{\top}$. The non-Gaussian part of $\ket{\psi_1}$   results only from the superposition of Fock states. 
The maximum Fock number  $n_{\text{max}}$ satisfies $n_{\text{max}} \le n_T$, where $n_T = n_2+n_3+ \cdots + n_N$ is the total number of detected photons. 
The inequality is saturated when $b_{1j} \ne 0$ for $j$ from 2 to $N$, which implies that the maximally supported non-Gaussian state is obtained when the unmeasured mode is fully connected with all other modes. 
The coefficients $\{c_n\}$ of Eq.~\eqref{eq:StateStructureSingle} are determined by \cite{su2019conversion}
\begin{eqnarray}\label{eq:Coefficients}
c_m c_n^* &\propto& 
\prod_{k=2}^{N} \bigg(\frac{\partial^2}{\partial \alpha_k \partial \beta_k^*} \bigg)^{n_k} \bigg[
\exp\bigg( \frac{1}{2} \vt{\gamma}_d^{\top} {\bf C} \vt{\gamma}_d + \vt{Y}^{\top} \vt{\gamma}_d \bigg) 
\nonumber\\
&&
\times
\bigg( \sum_{j=2}^N \kappa_j^* \alpha_j \bigg)^m \bigg(  \sum_{i=2}^N \kappa_i \beta_i^* \bigg)^n \bigg]
\bigg|_{\vt{\gamma}_d = 0},
\end{eqnarray}
where ${\bf C} = {\bf A} + {\bf R}_{dh} {\bf X}_2 {\bf R}_{hd} /(1 - |b_{11}|^2)$ and $\kappa_j = b_{1j}/ \sqrt{1 - |b_{11}|^2}$ for $j \ge 2$. 
Although Eq.~\eqref{eq:Coefficients} gives the product of two coefficients, it is easy to find $c_n/c_{n_{\text{max}}}$ from Eq.~\eqref{eq:Coefficients} 
and use the normalization condition to obtain a unique output state.

The measurement probability given a photon pattern $\vt{\bar{n}}$ can be computed as $P(\vt{\bar{n}}) = \bra{\vt{\bar{n}}}{\rm Tr}_h(\rho)\ket{\vt{\bar{n}}}$~\cite{PhysRevA.49.2993, PhysRevA.50.813, PhysRevLett.119.170501}, and one obtains \cite{su2019conversion}
\begin{align}\label{eq:ProbabilitySinlgemode}
&P(\bar{\vt{n}}) = \frac{\mathcal{ P}_0}{\bar{\vt{n}}!}
\bigg[\text{det} ({\bf I}_2 - {\bf X}_2 {\bf R}_{hh} ) \bigg]^{-1/2}
\exp \bigg( \frac{1}{2} \vt{y}_h^{\top} \vt{d} \bigg) \nonumber\\
&\times \prod_{k=2}^{N} \bigg(\frac{\partial^2}{\partial \alpha_k \partial \beta_k^*} \bigg)^{n_k}
\exp\bigg( \frac{1}{2} \vt{\gamma}_d^{\top} {\bf A}_p \vt{\gamma}_d + \vt{z}_p^{\top} \vt{\gamma}_d \bigg) \bigg|_{\vt{\gamma}_d = 0}, \nonumber\\
\end{align}
\begin{align}
&\mathcal{P}_0 = 2^{N} \bigg[ \text{det}\bigg( 2 {\bf V}^{(c)} + {\bf I}_{2N} \bigg) \bigg]^{-1/2} 
\exp\bigg( -\frac{1}{2} \vt{Q}^{(c) \top} \tilde{\vt{y}} \bigg),\nonumber \\ 
&{\bf A}_p = {\bf R}_{dd} + {\bf R}_{dh} ({\bf I}_2 - {\bf X}_2 {\bf R}_{hh} )^{-1} {\bf X}_2 {\bf R}_{hd},\nonumber\\ 
&\vt{z}_p = \vt{y}_d + {\bf R}_{dh} \vt{d}, ~~~~~~\bar{\vt{n}}! = n_2! n_3! \cdots n_N!.
\end{align}

\begin{table*}[tp]
\caption{ Preparation of an even cat state by detecting a two-mode Gaussian state with a PNRD. The even cat state is
approximated by $\hat{S}(\zeta_1) (c_0 \ket{0} + c_2 \ket{2})$. $\mathcal{F}_{\text{max}}$ is the maximum fidelity, $P_{\text{max}}$ is the maximum success probability,
$\zeta_{01}$ and $\zeta_{02}$ are the squeezing parameters of input squeezed vacuum states, and $\theta$ is the parameter of the beamsplitter defined as 
$e^{\theta(\hat{a}_1 \hat{a}_2^{\dag} - \hat{a}_1^{\dag} \hat{a}_2)}$.} 
\label{tab:ApproxEvenCat}
\centering
\begin{center}
\resizebox{0.85\textwidth}{!}{
    \begin{tabular}{| c | c | c | c | c | c | c | c |}
    \hline  \hline
    ~~~~~~$\alpha$~~~~~~ & ~~~~~~$\mathcal{F}_{\text{max}}$~~~~~~ & ~~~~~~$\zeta_1$~~~~~~& ~~~~~~$c_0/c_2$~~~~~~ &
   ~~~~~~$P_{\text{max}}$~~~~~~&~~~~~~$\zeta_{01}$~~~~~~&~~~~~~$\zeta_{02}$~~~~~~ &~~~~~~$\theta$~~~~~~ \\ \hline 
    0.25 & 1.0000 & 0.0115 & 27.717 & 18.12 \% & 1.1587 & $-0.0136$ & $-1.3965$ \\ \hline 
    0.50 & 1.0000 & 0.0458 & 6.9428 & 15.49 \% & 1.1936 & $-0.0499$ & $1.2351$ \\ \hline 
    0.75 & 0.9999 & 0.1025 & 3.1112 & 12.87 \% & 1.2447 & $-0.0982$ & $-1.0927$ \\ \hline 
    1.00 & 0.9999 & 0.1796 & 1.7885 & 11.20 \% & 1.3073 & $-0.1474$ & $-0.9686$ \\ \hline
    1.25 & 0.9991 & 0.2730 & 1.1932 & 10.55 \% & 1.3780 & $-0.1898$ & $0.8606$ \\ \hline
    1.50 & 0.9958 & 0.3763 & 0.8841 & 10.51 \% & 1.4546 & $-0.2228$ & $-0.7668$ \\ \hline
    1.75 & 0.9870 & 0.4832 & 0.7082 & 10.73 \% & 1.5346 & $-0.2464$ & $-0.6859$ \\ \hline
    2.00 & 0.9709 & 0.5884 & 0.6011 & 11.01 \% & 1.6150 & $-0.2626$ & $-0.6170$ \\ 
    \hline
    \end{tabular}
   }
\end{center}
\end{table*}

{\it {Number of independent coefficients}.} -- A natural question  arises as to how many of the $\{c_n\}$ in Eq.~\eqref{eq:Coefficients} are independent. This is crucial because it determines what states one can prepare and  also characterizes the extent of non-Gaussianity generated by a PNR measurement on a multimode Gaussian state. We observe by Eq.~\eqref{eq:Coefficients} that  when $b_{1 j} \ne 0$ for all $j$ from 2 to $N$, the maximal Fock number $n_{\text{max}}$ is equal to the total number of detected photons $ n_T$. In principle, there are no restrictions on $n_T$ but the number of independent $\{c_n\}$ is limited because there
is a finite number of complex parameters $N(2N+3)/2$ resulting from the covariance and mean of the pure Gaussian state. 
We find that the number of independent $\{c_n\}$ is smaller, and the redundant degrees of freedom
allow us to search for the optimal Gaussian state that maximizes the success probability of the output state. 

In the following, we will assume that $b_{1 j} \ne 0$ (or $\kappa_j \ne 0$) for all $j$ from 2 to $N$. 
By defining an $(N-1)$-component vector $\vt{\mu}$ as $\mu_j = Y_j/\kappa_j^*$, and a symmetric matrix ${\bf F}$ whose entries are 
$f_{ij} = b_{11}^* + \frac{b_{ij}}{\kappa_i \kappa_j}$, where $i, j = 2, 3, \cdots, N$, the ratio $c_n/c_{n_T}$ can now be written as \cite{su2019conversion}
\begin{align}\label{eq:CoeffRatio}
&\frac{c_n}{c_{n_T}} = 
\prod_{k=2}^{N} \bigg(\frac{\partial^2}{\partial \omega_k \partial \sigma_k^*} \bigg)^{n_k}
\exp(Z) \frac{ \mathcal{W}^n \, \Sigma^{n_T}}{ \sqrt{n! \, (n_T!)^3}}
\bigg|_{\vt{\omega} = \vt{\sigma} = 0}, 
\nonumber\\
&Z = \frac{1}{2} (\vt{\sigma}^{*}, \vt{\omega})^{\top} \, {\bf C}_{\text{rn}} \begin{pmatrix} \vt{\sigma}^{*} \\ \vt{\omega} \end{pmatrix} 
	+ (\vt{\mu}^{*}, \vt{\mu})^{\top} \begin{pmatrix} \vt{\sigma}^{*} \\ \vt{\omega} \end{pmatrix},\nonumber\\
&\mathcal{W}= \sum_{j=2}^N \omega_j,~~ \Sigma = \sum_{i=2}^N \sigma_i^*, ~ 	{\bf C}_{\text{rn}} = {\bf F} \oplus {\bf F}^*.
\end{align}

Therefore, the ratio $c_n/c_{n_T}$ is uniquely determined by the vector $\vt{\mu}$
and the matrix ${\bf F}$. Performing the partial derivatives in Eq.~\eqref{eq:CoeffRatio} results in a polynomial of 
$\mu_j$ and $f_{ij}$. The total number of independent complex parameters consisting of the components of $\vt{\mu}$ and the entries of {\bf F} (being symmetric)
is $ \mathfrak{D} = (N+2)(N-1)/2$. The problem of determining the number of independent $\{c_n\}$ can be formulated as follows. Suppose that $\mu_j$ and $f_{ij}$ are unknown
and have to be solved from $n_T$ nonlinear polynomial equations which come from Eq.~\eqref{eq:CoeffRatio} by taking $n = 0, 1, \cdots, n_T-1$. If $n_T < \mathfrak{D}$, 
the nonlinear equations are under-determined, which means that for a given set of $\{c_n\}$ there is an infinite number of solutions. This implies that there are many Gaussian states
that can generate the same non-Gaussian state. If $n_T > \mathfrak{D}$, the nonlinear equations are over-determined and there is no guarantee for the existence of
solutions for an arbitrary given set  $\{c_n\}$, which means they are not independent. The situation is subtle for the case of $n_T = \mathfrak{D}$. If there exists solutions,
the number of solutions is finite. It is also possible that there exists no solutions. We verified this for the $N=2,3$ cases via stochastic numerical simulations and found that when $n_T = \mathfrak{D}$ there
always exists a finite number of solutions, leading us to  the conjecture:
\begin{conjecture}
Measuring $(N-1)$ modes of an $N$-mode pure Gaussian state using PNR detectors outputs a coherent superposition of Fock states with at most $(N+2)(N-1)/2$ independent coefficients. 
\end{conjecture}
The above conjecture captures the power of PNRDs to  generate non-Gaussian states by measuring few modes of a Gaussian state. 
Equation~\eqref{eq:CoeffRatio} also provides a systematic way to generate target states. If the target state is of the form of Eq.~\eqref{eq:StateStructureSingle}, 
then it can be generated with fidelity one. Otherwise, one can approximate the target state 
using Eq.~\eqref{eq:StateStructureSingle}. A sufficiently high fidelity can always be obtained if $n_{\text{max}}$ is sufficiently large. 
The procedure to prepare a target state is: (1) approximate the target state using Eq.~\eqref{eq:StateStructureSingle} and find $\zeta$, $\beta$ and $c_n$; (2) solve the nonlinear equations obtained from Eq.~\eqref{eq:CoeffRatio}; (3) find ${\bf R}$ and $\vt{y}$ that optimize the success 
probability; (4) compute the covariance matrix ${\bf V}^{(c)}$ and the displacement vector $\vt{Q}^{(c)}$ of the measured Gaussian state. We now discuss a relevant examples of  non-Gaussian state generation.

{\it Schr\"odinger's cat states.} -- These  states are  superpositions of two coherent states with opposite phases:
$\ket{\text{cat}_{e/o}} \sim \ket{\alpha} \pm \ket{-\alpha}$, $e/o$ labeling even or odd. Bosonic codes based on cat states allow for 
fault-tolerant quantum computing~\cite{PhysRevA.68.042319, PhysRevLett.100.030503}. Here, we focus on the even cat state which is a 
superposition of only even Fock states. 
The even cat state can be well approximated by $c_0 \ket{0} + c_2 \ket{2}$ for small $\alpha$ and by $\hat{S}(\zeta_1) (c_0 \ket{0} + c_2 \ket{2})$ 
when $\alpha$ is big (see Table~\ref{tab:ApproxEvenCat}), where $\hat{S}(\zeta_1)$ is a squeezing operator with parameter $\zeta_1$. 
It is evident that $\hat{S}(\zeta_1) (c_0 \ket{0} + c_2 \ket{2})$ is in the form of Eq.~\eqref{eq:StateStructureSingle} and can be generated by detecting a pure two-mode 
Gaussian state using a PNRD and post selecting the measurement outcome with two photons. 

We summarize the results in Table~\ref{tab:ApproxEvenCat} which shows
the maximum success probability, the corresponding input squeezing and beamsplitter parameters. 
One can see that a high fidelity ($> 97\%$) and a high success probability ($>10\%$) can be achieved for $\alpha \le 2$, and the 
requirement for input squeezing is $ \zeta_{01} \in (1.1587, 1.6150$), i.e., $\sim 10-14$ dB. This  squeezing range is within 
current technology since $15$ dB squeezing has been demonstrated experimentally~\cite{PhysRevLett.117.110801}. It was demonstrated~\cite{PhysRevLett.120.073603} that the decoherence process can be 
substantially slowed down for squeezed cat states, which can  be generated using only offline squeezing in our framework. 

\begin{figure}
\scalebox{0.55}{\includegraphics{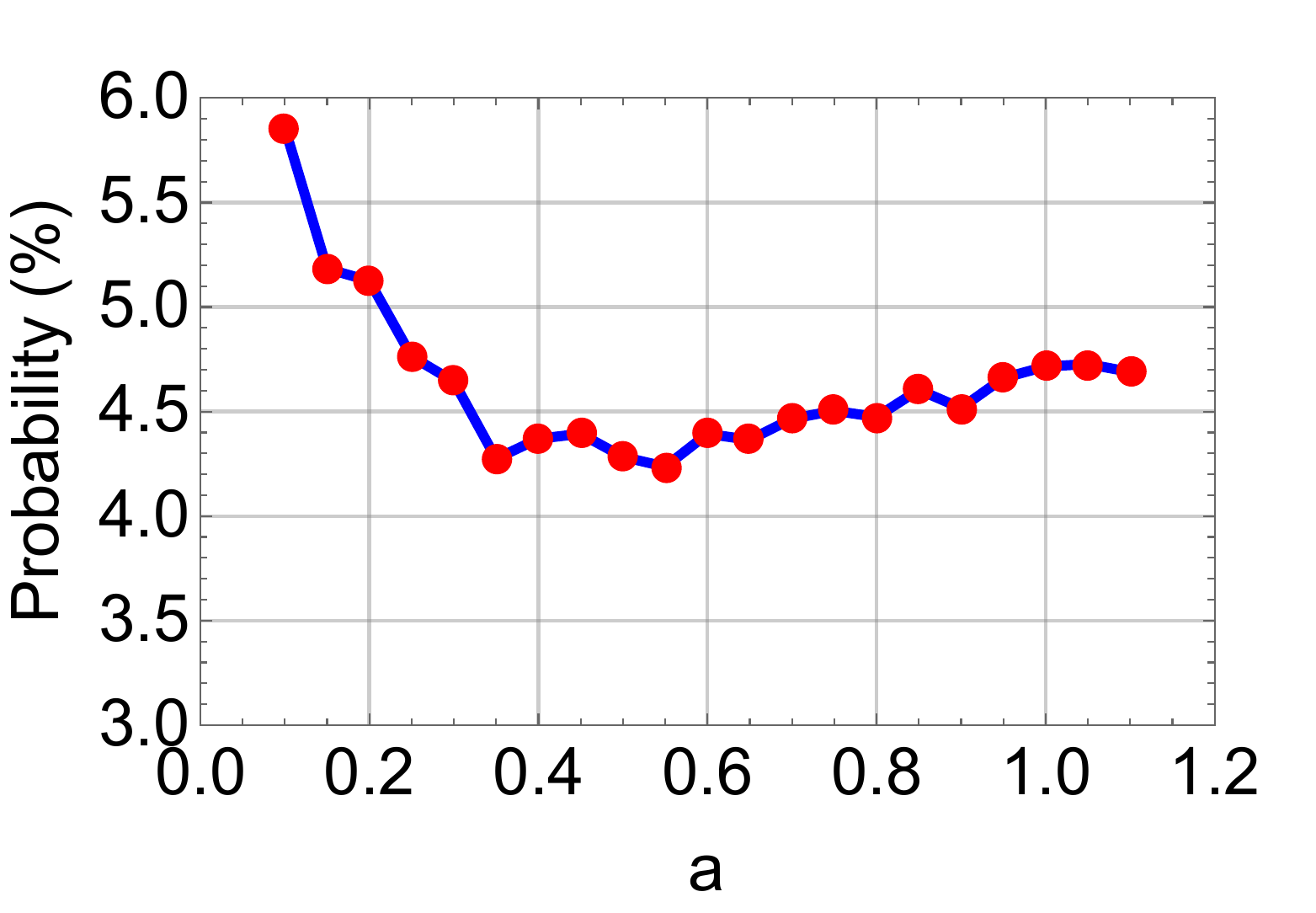}}
\caption{Graph showing the probability of producing $\ket{\phi_a}$ with perfect fidelity. A three mode circuit is used and the state is conditioned on detecting 1 photon and 2 photons at the second and third modes, respectively.} 
\label{fig:PlotONstate}
\end{figure}

{\it GKP states.} -- The Gottesman-Kitaev-Preskill (GKP) code was proposed to encode qubits in qumodes to protect against shifts errors in the  quadratures~\cite{PhysRevA.64.012310}
and photon loss~\cite{PhysRevA.97.032346}. However, generating the optical GKP codes is very challenging ~\cite{pirandola2004constructing, pirandola2006generating, vasconcelos2010all, PhysRevA.97.022341, PhysRevA.95.053819}. Here, we use the proposed formalism to conditionally generate an approximate GKP state 
$\psi_{\text{GKP}}(q; \Delta) = k_0 \sum_{s=-\infty}^{+\infty} \exp[{-2 \pi \Delta^2 s^2 - (q-2 \sqrt{\pi} s)^2/(2\Delta^2)}]$,
where $k_0=N_0(\pi \Delta^2)^{-1/4}$, $\Delta$ is the standard deviation and $N_0$ is the normalization. We use $\hat{S}(\zeta_1) (c_0 \ket{0} + c_2 \ket{2} + c_4 \ket{4} )$ to approximate a GKP state with $\Delta=0.35$, corresponding
to 9.12 dB of squeezing. The best fidelity $81.8\%$ is obtained with parameters: $\zeta_1 = 0.294, c_0 = 0.669, c_2 = -0.216, c_4 = 0.711$. We generate the approximate
state by measuring two modes of a three-mode Gaussian state and post select the photon number pattern $\bar{\vt{n}} = (2, 2)$. The best success probability we obtained
is  $\approx 1.1\%$. 

{\it {Weak cubic phase state}.} -- These states are represented as $\ket{\varphi}_a = (1+5|a|^{2}/2)^{-1/2}\left[\ket{0}+i a\sqrt{3/2}\ket{1}+i a \ket{3}\right]$, for real $a$. 
Such states can be combined in a gate-teleportation scheme to implement weak cubic phase gates on input states. A recent proposal ~\cite{sabapathy2018near} obtained optical schemes (which falls within our framework) using machine learning algorithms to generate these states with a success probability $1-2\%$, vastly improving previous techniques by four orders of magnitude. In Fig.~\ref{fig:PlotONstate}, we find using our techniques, that we improve the success probability to $4-6\%$ at the cost of increased squeezing requirements. 

{\it Multimode states.} -- The generalization of the framework to multimode output states is straightforward and the structure of the output states is similar to Eq.~\eqref{eq:WignerPure} \cite{su2019conversion}.
The procedure to find a target multimode non-Gaussian state is also similar to that of the single-mode non-Gaussian state.
We consider two examples of generating multimode non-Gaussian states: the $M$-mode W  state (denoted by $\mathbb{W}_M$) and NOON states \cite{su2019conversion}. 

The $\mathbb{W}_{M}$ state is an equal superposition of $\ket{{\bf 1}_k}$ for all possible $k$, where we define $\ket{{\bf 1}_k}$ as the state with one photon in the $k$-th mode
and zero photons in other modes. A $\mathbb{W}_{M}$ state can be generated by measuring one mode of an ($M+1$)-mode Gaussian state and post selecting 
the measurement outcome with one photon. We find that the $\mathbb{W}_M$ state can always be generated with fidelity $1$ and maximum success probability of $25\%$, which is independent of $M$. 

The NOON state is defined as $(\ket{N0} + \ket{0N})/\sqrt{2}$ with $N$ a positive integer. Generating NOON states via photon-number measurement on Gaussian 
states has been proposed~\cite{PhysRevA.97.053814}. 
Here, we use our formalism to generate NOON states with $N = 2, 3, 4$ by measuring multimode Gaussian states. 
The results are summarized in Table~\ref{tab:NG_apps}. The maximum success probability we obtained is significantly larger than that obtained in Fig.~4 of Ref.~\cite{PhysRevA.97.053814}. 

\begin{table}
\caption{An illustrative list of some non-Gaussian target states along with the fidelity (Fid.) to the state generated using our framework, the success probability (Prob.) comparerd to some  previous examples (Previous), and the number of total (T) and detected (D) modes of the pure Gaussian states that one begins with. }
\begin{tabular}{l|c|c|c|c}
\hline
\textbf{States} &{\bf Fid.} & {\bf Prob.} & ({\bf T, D})& {\bf Previous}\\
\hline
\hline
 $\ket{0} + \ket{2}$ & 1 & $10.48\%$ & (2, 1) & - \\
Cat  & $\sim 1.0$ & $\sim 10-20\%$ &(2, 1) & $7.5\%$ \cite{ourjoumtsev2007generation}\\
GKP & 0.818 & $1.1\%$ & (3, 2) & -\\
Weak cubic & 1 & $\sim 4-6\%$  & (3, 2) & $1\% -2\%$ \cite{sabapathy2018near} \\
\hline
$\mathbb{W}_{N}$  &1 & $25\%$ & (N+1, 1) & $ 1\%$ ~ \cite{PhysRevLett.120.130501}\\
NOON (N=2) &1 & $6.25\%$ & (4, 2) & $1.6\%$ ~\cite{PhysRevA.97.053814} \\
NOON (N=3) &1 & $1.54\%$ & (5, 3) & $0.19\%$ ~\cite{PhysRevA.97.053814} \\
NOON (N=4) &1 & $0.55\%$ & (6, 4)& $0.025\%$ ~\cite{PhysRevA.97.053814}\\
\hline
\end{tabular}
\label{tab:NG_apps}
\end{table}

{\it Conclusion.} -- We developed a detailed and systematic framework for the study of probabilistic generation of non-Gaussian states by measuring multimode Gaussian states via PNRDs. 
We derive analytic expressions for the output Wigner function and the measurement probability, which show explicitly the mapping between the properties of the multimode
Gaussian states and that of the heralded non-Gaussian states. The framework unifies many state preparation schemes, and more importantly, it provides a procedure to generate
a given target state with the best fidelity and success probability. We apply the proposed formalism to generate some important non-Gaussian states, and find that both the fidelity
and success probability are improved as compared to previous schemes. With the currently available PNRDs~\cite{magana2019multiphoton, tiedau2019scalability}, 
our framework would be a promising candidate to generate non-Gaussianity that is essential in applications like quantum metrology and fault-tolerant quantum computing using bosonic codes.

{\it Note.} -- During the completion of the work we were made aware of a related work \cite{gagatsos2019efficient}.

{\it {Acknowledgement}.} -- We thank Haoyu Qi, Kamil Br\'adler, Christian Weedbrook, Saikat Guha and Christos Gagatsos for insightful discussions.




\vspace{10 mm}

\bibliography{ref_short}

\end{document}